%Paper: hep-th/9208023
%From: "FARAGGI ALON" <FHALON@wiswic.weizmann.ac.il>
%Date: 9 Aug 92 14:22:00 WET

%
% Printing instructions:
%       This paper needs the macro packages phyzzx.tex and tables.tex
%       table 1. in \special{landscape} mode
%       should be stripped off and printed separately.
%
\input phyzzx
\tolerance=1000
\sequentialequations
\def\rl{\rightline}

\def\t1{{\tilde 1}}

\def\AEF{A.E. Faraggi}
\def\DVN{D. V. Nanopoulos}

\def\SSM{supersymmetric standard model}
\def\NPB#1#2#3{Nucl. Phys. B {\bf#1} (19#2) #3}
\def\PLB#1#2#3{Phys. Lett. B {\bf#1} (19#2) #3}

\def\MODA#1#2#3{Mod. Phys. Lett. A {\bf#1} (19#2) #3}
\def\IJMP#1#2#3{Int. J. Mod. Phys. A {\bf#1} (19#2) #3}

\REF\GSW{M. Green, J. Schwarz and E. Witten,
Superstring Theory, 2 vols., Cambridge
University Press, 1987.}
\REF\REVAMP{I. Antoniadis, J. Ellis,
J. Hagelin, and \DVN, \PLB{231}{89}{65}.}
\REF\FSU{T.T. Burwick, A.K. Kaiser and H.F. Muller \NPB{362}{91}{232};
A. Kagan and S. Samuel, CCNY--HEP--92/2; J. Lopez, D.V. Nanopoulos
and K. Yuan CTP--TAMU--92/11.}
\REF\SUTHREE{B. Greene {\it{el al.}},
Phys.Lett.{\bf B180} (1986) 69;
Nucl.Phys.{\bf B278} (1986) 667;
{\bf B292} (1987) 606;
R. Arnowitt and  P. Nath,
Phys.Rev.{\bf D39} (1989) 2006; {\bf D42}
(1990) 2498; Phys.Rev.Lett. {\bf 62} (1989) 222.}
\REF\SSM{L.E. Iba{\~n}ez {\it{et al.}}, Phys.Lett.
{\bf B191}(1987) 282; A. Font {\it{et al.}},
Phys.Lett. {\bf B210} (1988)
101; A. Font {\it{et al.}},
Nucl.Phys. {\bf B331} (1990)
421; D. Bailin, A. Love and S. Thomas,
Phys.Lett.{\bf B194} (1987) 385;
Nucl.Phys.{\bf B298} (1988) 75;
J.A. Casas, E.K. Katehou and C. Mu{\~n}oz,
Nucl.Phys.{\bf B317} (1989) 171.}
\REF\FNY{\AEF, D.V. Nanopoulos and K. Yuan, \NPB{335}{90}{437}.}
\REF\EU{\AEF, \PLB{278}{92}{131} .}
\REF\TOP{\AEF, \PLB{274}{92}{47}.}
\REF\SLM{\AEF, WIS--91/83/NOV--PH, WIS--92/16/FEB--PH ( To appear in Nucl.
Phys. B.). }
\REF\FFF{I. Antoniadis, C. Bachas, and C. Kounnas,
Nucl.Phys.{\bf B289}
(1987) 87; I. Antoniadis and C. Bachas,
Nucl.Phys.{\bf B298} (1988)
586; H. Kawai, D.C. Lewellen, and S.H.-H. Tye,
Phys.Rev.Lett.{\bf57} (1986)
1832; Phys.Rev.{\bf D34} (1986) 3794;
Nucl.Phys.{\bf B288} (1987) 1;
R. Bluhm, L. Dolan, and P. Goddard,
Nucl.Phys.{\bf B309} (1988) 330.}
\REF\FIQ{A. Font, L.E. Iba{\~n}ez and F. Quevedo,
 Phys.Lett.{\bf B228} (1989) 79.}
\REF\FN{A.E. Faraggi and D.V. Nanopoulos, \MODA{6}{91}{61}.}
\REF\DSW{M. Dine, N. Seiberg and E. Witten,
Nucl.Phys.{\bf B289} (1987) 585.}
\REF\ALR{I. Antoniadis, G. K. Leontaris and
J. Rizos, \PLB{245}{90}{161}.}
\REF\KLN{S. Kalara, J. Lopez and D.V. Nanopoulos,
\PLB{245}{91}{421}; \NPB{353}{91}{650}.}
\REF\naturalness{A.E. Faraggi and D.V. Nanopoulos,
Texas A \& M University preprint CTP--TAMU--78, ACT--15;
\AEF, Ph.D thesis, CTP--TAMU--20/91, ACT--31.}
\REF\ADS{J.J. Atick, L.J. Dixon and A. Sen, \NPB{292}{87}{109};
S. Cecotti, S. Ferrara and M. Villasante, \IJMP{2}{87}{1839}.}
\REF\PDB{\PLB{204}{88}{}.}
\REF\SFMM{J. Lopez and \DVN, \NPB{338}{90}{73}, \PLB{251}{90}73;
J. Rizos and K. Tamvakis, \PLB{251}{90}{369}; I. Antoniadis,
J. Rizos and K. Tamvakis, \PLB{278}{92}{257}.}

\singlespace
\rl{SLAC--PUB--5845}
\rl{WIS--92/48/JUN--PH}
\rl{SSCL--preprint--124}
\rl{June, 1992}
\rl{T}
\normalspace
\smallskip
\titlestyle{\bf{Aspects of Nonrenormalizable Terms in a Superstring
Derived Standard--like Model}}
\author{Alon E. Faraggi
{\footnote*{e--mail address: fhalon@weizmann.bitnet}}}
\smallskip
\centerline{Stanford Linear Accelerator Center}
\centerline{Stanford University, Stanford, California, 94305}
\centerline{and}
\centerline {Department of Physics, Weizmann Institute of Science}
\centerline {Rehovot 76100, Israel{\footnote\ddag{Permanent address}}}
\titlestyle{ABSTRACT}

I investigate the role of nonrenormalizable terms, up to order N=8, in a
superstring derived standard--like model. I argue that
nonrenormalizable terms restrict
the gauge symmetry,
at the Planck scale, to be
$SU(3)\times SU(2)\times U(1)_{B-L}\times U(1)_{T_{3_R}}$
rather than $SU(3)\times SU(2)\times U(1)_Y$.
I show that breaking the gauge symmetry directly to the Standard
Model leads to breaking of supersymmetry at the
Planck scale, or to dimension four, baryon and lepton violating,
operators. I show that if the gauge symmetry is broken directly
to the Standard Model the cubic level
solution to the F and D flatness constraints is violated by higher
order terms, while if $U(1)_{Z^\prime}$ remains unbroken at the
Planck scale, the cubic level solution is valid to all orders
of nonrenormalizable terms.
I discuss the Higgs and fermion mass spectrum.
I demonstrate that realistic, hierarchical, fermion mass spectrum
can be generated in this model.

\singlespace
\vskip 0.5cm
\endpage
\normalspace
\pagenumber 1

\centerline{\bf 1. Introduction}
Superstring theories [\GSW] are believed to provide a consistent
framework for the unification of all the known fundamental
interactions. The superstring unification scale is
at the Planck scale.
At the electroweak scale the Standard Model is in good
agreement with experimental observations.
However, the Standard Model, and point field
theories in general, leave many problems unresolved. Among them,
the origin of the number of generations, the origin of Yukawa couplings
and their hierarchy, quantum gravity, etc.
These problems find natural solutions in superstring theories.
Thus, an extremely
important task is to connect the superstring with the Standard  Model.

Two approaches can be pursued to  derive the Standard Model from the
superstring. One is to use a GUT
symmetry at an intermediate energy scale.
Many attempts have been made in this direction and most notable are
the flipped $SU(5)$  [\REVAMP,\FSU]
and the $SU(3)^3$ models [\SUTHREE]. The second approach is to derive
the Standard Model directly from the superstring without any
non--abelian gauge symmetry at an intermediate energy scale
[\SSM,\FNY,\EU,\TOP,\SLM]. In refs. [\EU,\TOP,\SLM]
realistic standard--like models were constructed in the free fermionic
formulation [\FFF], with the following properties:

\parindent=-15pt

1. Three and only three generations of chiral fermions. There are no
additional generations and mirror generations which presumably get
massive at a high scale. This property of the standard--like models
leads to an unambiguous identification
of the different generations.

2. The gauge group is ${SU(3)_C}\times{SU(2)_L}\times{U(1)_{B-L}}\times
{U(1)_{{T_3}_R}}\times U(1)^n\times{hidden}$. $n$ reduces to one or
zero after application of the
Dine--Seiberg--Witten (DSW) mechanism.
The
$U(1)_{Z^\prime}={1\over2}{U(1)_{B-L}}-{2\over3}{U(1)_{{T_3}_R}}$
combination may be broken at
the Planck scale, by the DSW mechanism.
If it remains unbroken down to low energies, it results in a
gauged mechanism to suppress proton decay from
dimension four operators [\FIQ,\FN].

3. There are enough scalar doublets and singlets
to break the symmetry in a realistic way and to
generate realistic fermion
mass hierarchy [\TOP,\SLM].

4. Proton decay from dimension four and dimension
five operators is suppressed
due to gauged $U(1)$ symmetries [\SLM].

5. These models suggest an explanation for
the top-bottom mass hierarchy. At the trilinear level
of the superpotential, only the top quark
gets a non vanishing mass term.
The mass terms for the bottom quark
and for the lighter quarks and leptons are
obtained from nonrenormalizable terms.
Thus, only the top quark mass is
characterized by the electroweak scale
and the masses of the lighter quarks and
leptons are naturally suppressed [\TOP,\SLM].
The top--bottom mass hierarchy is correlated with the
requirement of a supersymmetric vacuum
at the Planck scale [\EU,\TOP,\SLM].

\parindent=15pt
\smallskip

In this paper I examine the role of nonrenormalizable terms
in these models. For finiteness, I focus on the model of Ref. [\EU].
Nonrenormalizable terms are expected to play an important role in the low
energy phenomenology of these
models. I show that because of
nonrenormalizable terms the favored
observable gauge symmetry at the Planck
scale is
${SU(3)_C}\times{SU(2)_L}\times{U(1)_{B-L}}\times
{U(1)_{{T_3}_R}}$.  I show that in this
case the solution to the
cubic level F and D flatness
constraints is obeyed to all orders. In contrast
if the gauge symmetry
is broken directly to the Standard Model, at the Planck scale,
the cubic level
constraints are violated by higher order terms. Moreover,
I illustrate that breaking of the gauge symmetry directly to the Standard
Model may induce dimension four operators
which mediate rapid proton decay.
I suggest that these considerations
restrict the possible gauge symmetry
at the Planck scale to be
${SU(3)_C}\times{SU(2)_L}\times{U(1)_{B-L}}\times
{U(1)_{{T_3}_R}}$. Furthermore, they may
nessecitate the existence of an additional
neutral gauge boson at low energies, with
$U(1)_{Z^\prime}={1\over2}U(1)_{B-L}-{2\over3}U(1)_{T_{3_R}}$.
I discuss the Higgs and fermion mass
matrices in this model. I show that this model can generate
realistic, hierarchical fermion mass spectrum.

The paper is organized as follows. In section 2,
I review the model
and its symmetries. I discuss the rules for obtaining
the non vanishing nonrenormalizable terms and emphasize the
special properties of the standard--like
model which simplify the analysis.
In section 3,
I discuss the F and D flatness constraints.
In sections 4 and 5,
I discuss the implications of nonrenormalizable terms
on proton decay and on the fractionally
charged states. In sections
6 and 7,
I discuss the Higgs and fermion mass matrices. Section 8
concludes the paper.

\bigskip
\centerline{\bf 2. The superstring model}

The superstring model is constructed in the
free fermionic formulation [\FFF].
The model is generated by a basis of
eight boundary condition vectors.
The first five vectors in the
basis consist of the
NAHE{\footnote\dag
{This set was first constructed by Nanopoulos, Antoniadis,
Hagelin and Ellis, in the construction of the flipped $SU(5)$,
{\it nahe}
=pretty in Hebrew.}}
set, $\{{{\bf 1},S,b_1,b_2,b_3}\}$
[\REVAMP,\naturalness,\SLM].
This set is common to all the realistic models in the free
fermionic formulation [\REVAMP,\FNY,\ALR,\EU,\TOP,\SLM].
The important functions of the NAHE set are emphasized in
Ref. [\naturalness,\SLM]. The three vectors that extend the NAHE set
and the choice of generalized GSO coefficients
are given in table 1. The notation in the table emphasizes the
division of the internal fermions according to their
division by the NAHE set. In particular, it emphasizes the
division and assignment of boundary conditions to the set of real
fermions $\{y^i,\omega^i\vert{\bar y}^i,{\bar\omega}^i\}$
$(i=1,\cdots,6)$. The boundary conditions for this set of internal
fermions determine many of the properties of the low energy
spectrum [\SLM].

The gauge group after application of the generalized GSO
projections is

 Observable{\footnote*{$U(1)_C={1\over2}U(1)_{B-L}$,
         $U(1)_L={1\over2}U(1)_{T_{3_R}}$.}}
: $SU(3)_C\times U(1)_C\times SU(2)_L\times U(1)_L\times U(1)^6$

 Hidden{\footnote\#{Hidden here means that the states which are
identified with the chiral generations do not transform under the hidden
gauge group [\naturalness].}}
\hskip 0.4cm    :   $SU(5)_H\times SU(3)_H\times U(1)^2.$

The weak hypercharge is uniquely given by
$U(1)_Y={1\over3}U(1)_C+{1\over2}U(1)_L$. The orthogonal combination
is given by $U(1)_{Z^\prime}=U(1)_C-U(1)_L$.
In the observable sector there are six horizontal $U(1)$
symmetries.
The first three, $U(1)_j$ $(j=1,2,3)$, correspond to the right--moving
world--sheet currents ${\bar\eta}_1{\bar\eta}_1^*$,
${\bar\eta}_2{\bar\eta}_2^*$ and ${\bar\eta}_3{\bar\eta}_3^*$.
The last three, $U(1)_{r_{j+3}}$ $(j=1,2,3)$,
correspond to the right--moving
world--sheet currents,
${\bar y}^3{\bar y}^6$, ${\bar y}^1{\bar\omega}^5$ and
${\bar\omega}^2{\bar\omega}^4$, respectively.
For every right--moving $U(1)$ symmetry correspond
a left--moving global $U(1)$ symmetry. The first three
correspond to the charges of the supersymmetry generator
$\chi^{12}$, $\chi^{34}$ and $\chi^{56}$. The last three,
$U(1)_{\ell_{j+3}}$ $(j=1,2,3)$,
correspond to the complexified left--moving fermions
$y^3y^6$, $y^1\omega^5$ and $\omega^2\omega^4$.
Finally the model contains six Ising model sigma operators
which are obtained by pairing a left--moving
real fermion with a right--moving real fermion,
$\sigma^i_\pm=\{\omega^1{\bar\omega}^1,
y^2{\bar y}^2, \omega^3{\bar\omega}^3, y^4{\bar y}^4,
y^5{\bar y}^5, \omega^6{\bar\omega}^6\}_\pm$.

The full massless spectrum is analyzed by using a FORTRAN
program. The program takes as input the basis vectors
$B=\{b_1,\cdots,b_8\}$, and the GSO coefficients $c\left(\matrix{b_i\cr
b_j\cr}\right)$, $(i,j=1,\cdots,8)$. The program checks the modular
invariance rules, spans the additive group $\Xi=\sum_jn_jb_j$;
$(j=1,\cdots,8)$, selects the sectors in $\Xi$ which lead to massless
states and performs the GSO projections. It calculates the traces of the
of the $U(1)$ symmetries and evaluates the
quantum numbers of the massless states under all the symmetries in the
model. This output is read by subsequent programs which can analyze the
superpotential up to any order (the limit being a sensible CPU time
limit). This program enables a thorough exploration of a wider range
of models rather than specific isolated examples. Combined with the
conformal field theory techniques for evaluating corralators between
vertex operators, and the Renormalization Group Equations (RGE),
it provides powerful machinery for studying the phenomenology of the
superstring models.

The following massless states are produced by the sectors $b_{1,2,3}$,
$S+b_1+b_2+\alpha+\beta$, $O$ and their superpartners in the observable
sector:

(a) The massless spectrum contains three generations of chiral fermions
from the sectors $b_1$, $b_2$ and $b_3$:
$G_\alpha=e_{L_\alpha}^c+u_{L_\alpha}^c+N_{L_\alpha}^c+d_{L_\alpha}^c+
Q_\alpha+L_\alpha$ $(\alpha=1,\cdots,3)$, where
$$\eqalignno{{e_L^c}&\equiv [(1,{3\over2});(1,1)];{\hskip .6cm}
{u_L^c}\equiv [({\bar 3},-{1\over2});(1,-1)];{\hskip .2cm}
Q\equiv [(3,{1\over2});(2,0)]{\hskip 2cm}
&(1a,b,c)\cr
{N_L^c}&\equiv [(1,{3\over2});(1,-1)];{\hskip .2cm}
{d_L^c}\equiv [({\bar 3},-{1\over2});(1,1)];{\hskip .6cm}
L\equiv [(1,-{3\over2});(2,0)]{\hskip 2cm}
&(1d,e,f)\cr}$$
of $SU(3)_C\times U(1)_C\times SU(2)_L\times U(1)_L$,
with charges under the
six horizontal $U(1)$s.
 From the sector $b_1$ we obtain
$${\hskip .2cm}  ({e_L^c}+{u_L^c})_{{1\over2},0,0,{1\over2},0,0}+
({d_L^c}+{N_L^c})_{{1\over2},0,0,{-{1\over2}},0,0}+
(L)_{{1\over2},0,0,{1\over2},0,0}+(Q)_{{1\over2},0,0,-{1\over2},0,0},
\eqno(2a)$$ from the sector $b_2$
$${\hskip .2cm} ({e_L^c}+{u_L^c})_{0,{1\over2},0,0,{1\over2},0}+
({N_L^c}+{d_L^c})_{0,{1\over2},0,0,-{1\over2},0}+
(L)_{0,{1\over2},0,0,{1\over2},0}+
(Q)_{0,{1\over2},0,0,-{1\over2},0},\eqno(2b)$$ and from the sector $b_3$
$$ {\hskip .2cm} ({e_L^c}+{u_L^c})_{0,0,{1\over2},0,0,{1\over2}}+
({N_L^c}+{d_L^c})_{0,0,{1\over2},0,0,-{1\over2}}+
(L)_{0,0,{1\over2},0,0,{1\over2}}+(Q)_{0,0,{1\over2},0,0,-{1\over2}}.
\eqno(2c)$$
The vectors $b_1$, $b_2$ and $b_3$ are the only vectors in  the
additive group $\Xi$ that produce spinorial $16$ of $SO(10)$.
This is in contrast to the case in which the $SO(10)$
symmetry is broken to $SU(5)\times U(1)$ [\REVAMP]
or to $SO(6)\times SO(4)$ [\ALR]. There the massless spectrum
contains additional $16$ and ${\bar{16}}$ multiplets. The fact
that there are exactly three generations,
without any extra generations and
mirror generations, is unique to the choice of
$SU(3)\times SU(2)\times U(1)_C\times U(1)_L$ as the observable
gauge symmetry at the level of the spin structure.
This property of the standard--like models leads to an unambiguous
identification of the hierarchical generations.

(b) The ${S+b_1+b_2+\alpha+\beta}$ sector gives
$$\eqalignno{h_{45}&\equiv{[(1,0);(2,1)]}_
{-{1\over2},-{1\over2},0,0,0,0} {\hskip .5cm}
D_{45}\equiv{[(3,-1);(1,0)]}_
{-{1\over2},-{1\over2},0,0,0,0}&(3a,b)\cr
\Phi_{45}&\equiv{[(1,0);(1,0)]}_
{-{1\over2},-{1\over2},-1,0,0,0}  {\hskip .5cm}
\Phi^{\pm}_1\equiv{[(1,0);(1,0)]}_
{-{1\over2},{1\over2},0,\pm1,0,0}&(3c,d)\cr
\Phi^{\pm}_2&\equiv{[(1,0);(1,0)]}_
{-{1\over2},{1\over2},0,0,\pm1,0} {\hskip .5cm}
\Phi^{\pm}_3\equiv{[(1,0);(1,0)]}_
{-{1\over2},{1\over2},0,0,0,\pm1}&(3e,f)\cr}$$
(and their conjugates ${\bar h}_{45}$, etc.).
The states are obtained by acting on the vacuum
with the fermionic oscillators
${\bar\psi}^{4,5},{\bar\psi}^{1,...,3},{\bar\eta}^3,{\bar y}^3\pm
i{\bar y}^6,{\bar y}^1\pm{i{\bar\omega}^5},
{\bar\omega}^2{\pm}i{\bar\omega}^4$,
respectively  (and their complex conjugates for ${\bar h}_{45}$, etc.).

(c) The Neveu--Schwarz $O$ sector gives, in addition to  the graviton,
dilaton, antisymmetric tensor and spin 1 gauge bosons,  the
following scalar representations:

Electroweak doublets and singlets:
$$\eqalignno{{h_1}&\equiv{[(1,0);(2,-1)]}_{1,0,0,0,0,0}
{\hskip 2cm}\Phi_{23}\equiv{[(1,0);(1,0)]}_{0,1,-1,0,0,0}&(4a)\cr
{h_2}&\equiv{[(1,0);(2,-1)]}_{0,1,0,0,0,0}
{\hskip 2cm}\Phi_{13}\equiv{[(1,0);(1,0)]}_{1,0,-1,0,0,0}&(4b)\cr
{h_3}&\equiv{[(1,0);(2,-1)]}_{0,0,1,0,0,0}
{\hskip 2cm}\Phi_{12}\equiv{[(1,0);(1,0)]}_{1,-1,0,0,0,0}&(4c)\cr}$$
(and their conjugates ${\bar h}_1$, etc.).
Finally, the Neveu--Schwarz sector gives rise to three singlet
states that are neutral under all the U(1) symmetries.
$\xi_{1,2,3}:{\hskip .2cm}{\chi^{12}_{1\over2}{\bar\omega}^3_{1\over2}
{\bar\omega}^6_{1\over2}{\vert 0\rangle}_0},$
 ${\chi^{34}_{{1\over2}}{\bar y}_{1\over2}^5{\bar\omega}_{1\over2}^1
{\vert 0\rangle}_0},$
 $\chi^{56}_{1\over2}{\bar y}_{1\over2}^2{\bar y}_{1\over2}^4
{\vert 0\rangle}_0.$

The sectors $b_i+2\gamma+(I){\hskip .2cm} (i=1,..,3)$ give vector
representations which are
$SU(3)_C\times SU(2)_L\times {U(1)_L}\times {U(1)_C}$
singlets (see Table 1). The vectors with some combination
of $(b_1,b_2,b_3,\alpha,\beta)$
plus $\gamma+(I)$ (see Table 2) give  representations which transform
under $SU(3)_C\times SU(2)_L\times {U(1)_L}\times {U(1)_C}$, most of
them singlets, but carry either
$U(1)_Y$ or $U(1)_{Z^\prime}$ charges. Some of these states carry
fractional charges $\pm{1\over2}$ or $\pm{1\over3}$.
There are no representations that transform nontrivially both under the
observable
and hidden sectors. The only mixing
which occurs is of states that transform
nontrivially under the observable or
hidden sectors and carry U(1) charges
under the hidden or observable sectors, respectively.

The non vanishing trilevel
terms in the superpotential of the model are
$$\eqalignno{W&=\{(
{u_{L_1}^c}Q_1{\bar h}_1+{N_{L_1}^c}L_1{\bar h}_1+
{u_{L_2}^c}Q_2{\bar h}_2+{N_{L_2}^c}L_2{\bar h}_2+
{u_{L_3}^c}Q_3{\bar h}_3+{N_{L_3}^c}L_3{\bar h}_3)
+{{h_1}{\bar h}_2{\bar\Phi}_{12}}
+{h_1}{\bar h}_3{\bar\Phi}_{13}\cr
&\qquad
+{h_2}{\bar h}_3{\bar\Phi}_{23}
+{\bar h}_1{h_2}{\Phi_{12}}
+{\bar h}_1{h_3}{\Phi_{13}}
+{\bar h}_2{h_3}{\Phi_{23}}
+\Phi_{23}{\bar\Phi}_{13}{\Phi}_{12}
+{\bar\Phi}_{23}{\Phi}_{13}{\bar\Phi}_{12}
+{\bar\Phi}_{12}({\bar\Phi}_1^+{\bar\Phi}_1^-\cr
&\qquad
+{\bar\Phi}_2^+{\bar\Phi}_2^-
+{\bar\Phi}_3^+{\bar\Phi}_3^-)
+{\Phi_{12}}(\Phi_1^-\Phi_1^+
+\Phi_2^-\Phi_2^+
+\Phi_3^-\Phi_3^+)+{1\over2}\xi_3(\Phi_{45}{\bar\Phi}_{45}
+h_{45}{\bar h}_{45}+D_{45}{\bar D}_{45}\cr
&\qquad+\Phi_1^+{\bar\Phi}_1^++
\Phi_1^-{\bar\Phi}_1^-+\Phi_2^+{\bar\Phi}_2^++\Phi_2^-{\bar\Phi}_2^-
+\Phi_3^+{\bar\Phi}_3^++\Phi_3^-{\bar\Phi}_3^-)
+h_3{\bar h}_{45}\Phi_{45}+{\bar h}_3h_{45}{\bar\Phi}_{45}\}\cr
&\qquad
+\{{1\over2}[\xi_1(H_{19}H_{20}+
H_{21}H_{22}+H_{23}H_{24}+H_{25}H_{26})
+\xi_2(H_{13}H_{14}+H_{15}H_{16}+H_{17}H_{18})]\cr
&\qquad
+{\bar\Phi}_{23}H_{24}H_{25}
+{\Phi}_{23}H_{23}H_{26}+h_2H_{16}H_{17}
+{\bar h}_2H_{15}H_{18}
+{e_{L_1}^c}{H_{10}}{H_{27}}
+{e_{L_2}^c}{H_8}{H_{29}}
+({V_1}{H_9}\cr
&\qquad
+{V_2}{H_{11}}){H_{27}}+{V_6}{H_5}{H_{29}}
+{{\bar\Phi}_{45}}{H_{17}}{H_{24}}
+{D_{45}}{H_{18}}{H_{21}}
+{h_{45}}{H_{16}}{H_{25}}\}\quad&(5)\cr}$$
where a common normalization constant ${\sqrt 2}g$ is assumed.

Nonrenormalizable contributions to the superpotential are obtained
by calculating corralators between vertex operators
$$A_N\sim\langle V_1^fV_2^fV_3^b\cdot\cdot\cdot V_N^b\rangle,\eqno(6)$$
where $V_i^f$ $(V_i^b)$ are the fermionic (scalar)
components of the vertex operators.
The non vanishing terms are obtained by
applying the rules of Ref. [\KLN].
To obtain the correct ghost charge some of the
vertex operators are picture
changed by taking
$$V_{q+1}(z)=\lim_{w\to z}exp(c)(w)T_F(w)V_{q}(z),\eqno(7)$$
where $T_F$ is the super current and in the
fermionic construction is given by
$$T_F=\psi^\mu\partial_\mu X+i{\sum_{I=1}^6}\chi_{_I}{y_{_I}}
\omega_{_I}=T_F^0+T_F^{-1}+T_F^{+1}\eqno(8)$$
with
$$T_F^{-1}=e^{-i\chi^{12}}\tau_{_{12}}+e^{-i\chi^{34}}\tau_{_{34}}
+e^{-i\chi^{56}}\tau_{_{56}}{\hskip .5cm};
{\hskip .5cm}T_F^{-1}=(T_F^{+1})^*\eqno(9)$$
where
$\tau_{_{ij}}={i\over{\sqrt2}}(y^i\omega^i+y^j\omega^j)$
and $e^{\chi^{ij}}={1\over\sqrt{2}}(\chi^i+i\chi^j)$.

Several observations simplify the analysis of the
potential non vanishing
terms. First, it is observed that only the $T_F^{+1}$ piece of $T_F$
contributes to $A_N$ [\KLN]. Second, in the standard--like model
the pairing of left--moving fermions is
$y^1\omega^5$, $\omega^2\omega^4$ and $y^3y^6$.
One of the fermionic states in every term $y^i\omega^i$ $(i=1,...,6)$
is
complexified and therefore can be written,
for example for $y^3$ and $y^6$,
as
$$y^3={1\over{\sqrt2}}(e^{iy^3y^6}+e^{-iy^3y^6}),
y^6={1\over{\sqrt2}}(e^{iy^3y^6}-e^{-iy^3y^6}).\eqno(10)$$
Consequently, every picture changing operation changes the
total
$U(1)_\ell=U(1)_{\ell_4}+U(1)_{\ell_5}+U(1)_{\ell_6}$ charge
by $\pm1$. An odd (even)  order term
requires an even (odd) number of picture changing
operations to get the correct ghost number [\KLN].
Thus, for $A_N$ to be non vanishing,
the total $U(1)_\ell$ charge, before picture
changing, has to be an odd (even)
number, for even (odd) order terms, respectively.
Similarly, in every pair $y_i\omega_i$, one real
fermion, either $y_i$ or
$\omega_i$, remains real and is paired with the corresponding
right--moving real fermion to produce an Ising model sigma operator.
Every picture changing operation changes the number of left--moving
real fermions by one.
This property of the standard--like model significantly
reduces the number of
potential non vanishing terms.

\bigskip
\centerline{\bf 3. F and D constraints}

The massless spectrum of the superstring model
contains six anomalous $U(1)$
symmetries. Of the six anomalous $U(1)$s only five can be rotated
by an orthogonal transformation and one combination remains
anomalous. The six combinations can be taken as [\EU]
$$\eqalignno{{U^\prime}_1&=U_1-U_2{\hskip .5cm},{\hskip .5cm}
{U^\prime}_2=U_1+U_2-2U_3,&(11a,b)\cr
{U^\prime}_3&=U_4-U_5{\hskip .5cm},{\hskip .5cm}
{U^\prime}_4=U_4+U_5-2U_6,&(11c,d)\cr
{U^\prime}_5&=U_1+U_2+U_3+2U_4+2U_5+2U_6,&(11e)\cr
U_A&=2U_1+2U_2+2U_3-U_4-U_5-U_6,&(11f)\cr}$$
with $Tr(Q_A)=180.$

The anomalous $U(1)$ generates a Fayet--Iliopoulos D--term by the VEV
of the dilaton field. Such a D--term, in general, breaks supersymmetry.
Supersymmetry is restored if there exist a
direction in the scalar potential
$\phi=\sum_i{\alpha_i\phi_i}$  which is F flat
and also D flat with respect to the non anomalous gauge symmetries
and in which
$\sum_i{Q_i^A{\vert\alpha_i\vert}^2}< 0$.
If such a direction exists, it
will acquire a VEV, canceling the
anomalous D--term, restoring supersymmetry
and stabilizing the vacuum [\ADS]. Since the fields
corresponding to such a flat
direction typically also carry charges for the non anomalous D--terms,
 a non trivial set of constraints
 on the possible choices of VEVs is imposed.
 It is, in general, a non trivial
problem to find solutions to the set of constraints.

The set of constraints is summarized in the following set of equations,
$$\eqalignno{{D_A}&={\sum _{k}}{Q_k^A}{\vert\chi_k\vert}^2=
{-g^2e^{\phi_D}\over{192\pi^2}}{Tr(Q_A)}&(12a)\cr
D_j'&=\sum_{k}{Q'}_k^j\vert\chi_k\vert^2=0{\hskip .3cm}
 j=1\cdots5&(12b)\cr
D_j&=\sum_{k}Q_k^j\vert\chi_k\vert^2=0{\hskip .3cm} j=C,L,7,8&(12c)\cr
W&={{\partial W}\over{\partial{\eta_i}}}=0&(12d)\cr}$$
where $\chi_k$ are the fields that get a
VEV and $Q_k^j$ is their charge
under the $U(1)_j$ symmetry. The set
$\{{\eta}_i\}$ is the set of fields with vanishing
VEV.

In the standard--like models
the solutions to the set of F and D
constraints divide into two kinds of
solutions.
Solutions which break $U(1)_{Z'}$ and those which do not. Only the
Neveu--Schwarz sector and the $b_1+b_2+\alpha+\beta$
sector produce $SO(10)$
singlets with negative  $Q_A$. Therefore, only
these sectors contribute to solutions
which keep both $U(1)_Y$ and $U(1)_{Z'}$  unbroken at the Plank scale.
For solutions which break $U(1)_{Z^\prime}$,
the states from the sectors $b_{1,2}+b_3+
\alpha+\gamma\pm(I)$, and the states
$\{N_1,N_2,N_3\}$ from the sectors $b_1$, $b_2$ and $b_3$,
can obtain a VEV as well.
These states have vanishing weak hypercharge but non vanishing
$U(1)_{Z'}$ charge.

The F flatness conditions derived from the cubic superpotential are
$$\eqalignno{
{{\bar\Phi}_{13}}{\Phi_{12}}&+{H_{23}}{H_{26}}=
{{\Phi}_{13}}{\bar\Phi_{12}}
+{H_{24}}{H_{25}}={{\Phi}_{23}}{\Phi_{12}}={{\bar\Phi}_{23}}
{\bar\Phi_{12}}=0&(13a)\cr
{{\bar\Phi}_{23}}{\Phi_{13}}&+{\bar\Phi}^+_i{\bar\Phi}^-_i=0&(13b)\cr
{\bar\Phi}_i^+{\bar\Phi}_{12}&+\Phi_i^-\xi_3=0&(13c)\cr
{\bar\Phi}_i^-{\bar\Phi}_{12}&+\Phi_i^+\xi_3=0&(13d)\cr
\Phi_{45}{\bar\Phi}_{45}&+\Phi_i^+{\bar\Phi}_i^+
+\Phi_i^-{\bar\Phi}_i^-=0&(13e)\cr
\Phi_{45}\xi_3&+H_{17}H_{24}=0&(13f)\cr
&\bar\Phi_{45}\xi_3=0&(13g)\cr
H_{19}H_{20}&+H_{23}H_{24}+H_{25}H_{26}=0&(13h)\cr
H_{13}H_{14}&+H_{17}H_{18}=0.&(13i)\cr}$$
For equations $(13b)-(13d)$ the barred equations have to be taken
as well.
In addition to these equations we have 12 constraints of the form
$H\xi$. The total number of F flatness constraints results
in 35 equations.

I focus first on solutions which do not break
$U(1)_{Z^\prime}$.
I show that in this case the cubic level solution is obeyed to all
orders of nonrenormalizable terms.  I demonstrate that solutions
which break $U(1)_{Z^\prime}$ do not hold to all orders.

For solutions which do not break
$U(1)_{Z^\prime}$, $\langle{H}\rangle=0$.
Therefore, the choice
$$\langle{\Phi_{12},{\bar\Phi}_{12},\xi_3}\rangle=0,\eqno(14)$$
satisfies the cubic level F constraints. I also impose
$\langle\Phi_{23},{\bar\Phi}_{45}\rangle=0$.
 In this case the set of cubic level F constraints
reduces to
$$\eqalignno{
{{\partial W}\over{\partial{\Phi_{12}}}}&=
{{\bar\Phi}_{23}}{\Phi_{13}}+{\bar\Phi}^+_i{\bar\Phi}^-_i=0&(15a)\cr
{{\partial W}\over{\partial{{\bar\Phi}_{12}}}}&=
{{\bar\Phi}_{13}}{\Phi_{23}}+{\Phi}^+_i{\Phi}^-_i=
{\Phi}^+_i{\Phi}^-_i=0&(15b)\cr
{{\partial W}\over{\partial{\xi_{3}}}}&=
\Phi_{45}{\bar\Phi}_{45}+\Phi_i^
+{\bar\Phi}_i^++\Phi_i^-{\bar\Phi}_i^-=0
&(15c)\cr}$$
where $W$ is the cubic superpotential and
summation on repeated indices is implied.

I now turn to discuss the implication of nonrenormalizable terms
on the cubic level
F flatness constraints. The order $N$ terms that have to be
investigated are of the form
$$\langle(\alpha\beta)^j(NS)^{N-j}\rangle{\hskip2cm}
(j=4,\cdots,N)\eqno(16)$$
where $(NS)$ denotes fields which belong to the Neveu--Schwarz sector and
$(\alpha\beta)$ denotes fields that belong to the sector
$b_1+b_2+\alpha+\beta$.
Without loss of generality we can choose two of the $(\alpha\beta)$
fields to be the two space--time
fermions in these corralators. The
$N=2$ world--sheet global $U(1)$ charges,
$(\chi_{_{12}},\chi_{_{34}},\chi_{_{56}})$,
for the $(\alpha\beta)$ fields
are $(0,0,{1\over2})$ for fermions and $(-{1\over2},-{1\over2},0)$
for scalars. All the Neveu--Schwarz fields in Eq. (16)
are scalar fields,
with charges $\chi_{_{ij}}=0$ or $-1$.
Of the Neveu--Schwarz singlets, only
$\Phi_{12}$, ${\bar\Phi}_{12}$ and $\xi_3$ carry
$U(1)_{\ell_3}$ charges.
We can always choose
a basis in which the $\chi_{_{56}}$ charge of these fields is picture
changed to zero. The picture changing operation on the $(\alpha\beta)$
scalars can only change them to $(\pm{1\over2},\pm{1\over2},0)$.
Therefore, all the terms of the form of Eq. (16) are not invariant
under $U(1)_{\ell_3}$. The conclusion is that all these terms vanish
identically to all orders. Thus, in models with
$SU(3)_C\times SU(2)_L\times U(1)_{B-L}\times U(1)_{T_{3_R}}$
gauge symmetry at the Planck scale the cubic level F flatness
solution is valid to all orders of nonrenormalizable terms.

I now turn to show that in models with broken $U(1)_{Z^\prime}$,
at the Planck scale, the cubic level solution is violated by higher
order terms. As an illustrative example I take the solution that was
found in Ref. [\EU]. With the set of non vanishing VEVs,
$\{H_{23},H_{18},{\bar\Phi}_{13},\Phi_{45},{\bar\Phi}_{23},
\Phi^+_2,{\bar\Phi}_3^-\}$, Eqs. (12) have the solution,
$${\vert{H_{23}}\vert}^2={\vert{H_{18}}\vert}^2=
{1\over3}{\vert{\Phi}_{45}\vert}^2={3\over2}{\vert{\bar\Phi}_{13}
\vert}^2={3\over2}{\vert{\bar\Phi}_{23}\vert}^2=
{1\over2}{\vert\Phi^+_2\vert}^2=
{\vert{\bar\Phi}_3^-\vert}^2={{g^2}\over
{16\pi^2}}.\eqno(18)$$ This set breaks the observable gauge
symmetry to
$SU(3)_C\times SU(2)_L\times {U(1)_Y}$.
This solution obeys the cubic level F and D flatness constraints.
At order seven we find the following non vanishing term,
$$H_{23}^2H_{18}^2\Phi_{45}^2\xi_2.$$
Thus the cubic level constraint
${{\partial W}\over{\partial{\xi_{2}}}}\equiv0$
is violated. Moreover, if $\xi_2$
gets a Planck scale VEV the
superpotential receives a contribution
of $O({M_{Pl}})$ and $W\not=0$.
Therefore, in models with broken $U(1)_{Z^\prime}$
the cubic level solution is violated
by higher order terms, while in models
with unbroken $U(1)_{Z^\prime}$
the cubic level F flatness solution is valid
to all orders. I would like to emphasize that giving a VEV to any
pair of singlets from the sectors $b_{1,2}+b_3+\alpha+\gamma\pm(I)$
leads to a violation of the cubic level F flatness solution at the
quintic or $N=7$ orders.
We could contemplate giving a VEV to one of the three
Standard Model singlets in the $16$ of $SO(10)$, $N_1$, $N_2$ or $N_3$
and, for example, to
$H_{23}$. In this case F violating terms do not appear up to $N=7$,
but may appear at orders higher than $N=7$.
However, as I show in the next section giving a Planck scale VEV to
$N_1$, $N_2$ or $N_3$
leads to problems with proton decay.

The number of flat directions is larger than the number of constraints.
Therefore, the solution to the F and D constraints is not unique.
However, once a specific choice has been made, the
phenomenology of the model is determined. In what follows bellow
I focus on one illustrative example.
An explicit solution which satisfies all the F and D constraints is
given by the following set of non vanishing VEVs
$$\{\Phi_{45},\Phi_{1,2}^\pm,{\bar\Phi}_{1,2,3}^-,\bar\Phi_1^+,
{\bar\Phi}_{23},{\bar\Phi}_{13},\Phi_{13},\xi_1,\xi_2\}\eqno(19)$$
with
$$\eqalignno{&{1\over3}\vert\Phi_{45}\vert^2
=\vert\Phi_{23}\vert^2=
{4\over3}\vert\Phi_{13}\vert^2={4\over3}\vert
{\bar\Phi}_{13}\vert^2={{g^2}\over{16\pi^2}}&(20a)\cr
&{1\over2}\vert{\bar\Phi}_1^-\vert^2=
\vert{\bar\Phi}_1^+\vert^2=\vert\Phi_2^+\vert^2=
{1\over2}\vert\Phi_2^-\vert^2=
\vert\Phi_3^-\vert^2={{g^2}\over{16\pi^2}}&(20b)\cr
&\vert\Phi_1^+\vert^2=\vert\Phi_1^-\vert^2=
({7\over8})^{1\over2}{{g^2}\over{16\pi^2}}&(20c)\cr
&\vert{\bar\Phi}_2^-\vert^2={\sqrt{7\over2}}{{1+{\sqrt{2}}}\over{60}}
{{g^2}\over{16\pi^2}}.
&(20d)\cr}$$

\bigskip
\centerline{\bf 4.  Dimension four operators }

In this section I show that nonrenormalizable terms induce effective
dimension four operators which may result in rapid proton decay.
It is well known that the most general supersymmetric standard model
gives rise to dimension four operators, which induce
rapid proton decay,
$${\eta_1}{u_{L}^C}{d_{L}^C}{d_{L}^C}+
{\eta_2}{d_{L}^C}QL$$
where generations indices are suppressed.
If $\eta_1,\eta_2$ are of $O(1)$,
the proton will decay instantly.
These dimension four operators are forbidden if the gauge symmetry
of the Standard Model is extended  by an additional $U(1)$ gauge symmetry
which is a combination of $B-L$, baryon number minus lepton
number, and $T_{3_R}$ [2]. This $U(1)$ symmetry is exactly the
$U(1)_{Z^\prime}$ which is derived in the superstring
standard--like models. The dimension four operators may still appear
from the nonrenormalizable terms,
$${\eta_1}({u_{L}^C}{d_{L}^C}{d_{L}^C}N_L^C)\Phi+
{\eta_2}({d_{L}^C}QLN_L^C)\Phi$$
where $\Phi$ is a combination of fields
that fixes the string selection rules
[3] and gets a VEV of $O(m_{pl})$, and
${N_L^C}$ is the Standard Model singlet
in the $16$ of $SO(10)$. Thus, the ratio
${{\langle{N_L^C}\rangle}\over{M_{Pl}}}$
controls the rate of proton decay. In the standard--like model, the
following non vanishing terms appear at order $N=6$,

$$\eqalignno{&(u_3d_3+Q_3L_3)d_2N_2\Phi_{45}
{\bar\Phi}_2^{-}&(21a)\cr
      +&(u_3d_3+Q_3L_3)d_1N_1\Phi_{45}\Phi_1^{+}&(21b)\cr
      +&u_3d_2d_2N_3\Phi_{45}{\bar\Phi}_2^{-}+
  u_3d_1d_1N_3\Phi_{45}\Phi_1^{+}&(21c)\cr
      +&Q_3L_1d_3N_1\Phi_{45}\Phi_3^+
  +Q_3L_1d_1N_3\Phi_{45}\Phi_3^+&(21d)\cr
      +&Q_3L_2d_3N_2\Phi_{45}{\bar\Phi}_3^-
  +Q_3L_2d_2N_3\Phi_{45}{\bar\Phi}_3^-.&(21e)\cr}$$

In section 7, I will show that the states in $G_3$ have to
be identified with the lightest generation. From Eqs. (20) and (21)
it is evident
that if any of
$N_1$, $N_2$ or $N_3$ gets a Planck scale VEV, dimension
four operators are induced,
which result in rapid proton decay. Thus,
we conclude that
$\langle{N_1,N_2,N_3}\rangle\equiv0$ at the Planck scale.
Moreover, since the coefficients in front of the terms in Eqs. (21)
are expected to be of order one [\KLN], a possible VEV for
$N_{1,2,3}$ has to be well
below the GUT scale.
This result, combined with the result of the previous section,
show that models with an $SU(3)\times SU(2)\times U(1)_{B-L}\times
U(1)_{T_{3_R}}$ observable gauge symmetry, at the Planck scale, are
favored over models with $SU(3)\times SU(2)\times U(1)_Y$.

\bigskip
\centerline{\bf 5.  Fractionally charged states }

The massless spectrum of the superstring model contains the following
singlet states with fractional charge $\pm{1\over2}$,
$$H_3,H_4,H_7,H_8,H_{11},H_{12},H_{29},H_{30}.\eqno(22)$$
These states do not transform under any of the non abelian
gauge groups in the model. Therefore, they are not confined by any non
abelian gauge symmetry. While many experimental searches for
fractional charges have been conducted, no reported observation of a
fractionally charged state has ever been confirmed and there are upper
bounds on the abundance of any such particle in the range of
$10^{-19}$ to $10^{-26}$ [\PDB]
of the nucleon abundance for charges between
$1\over3$ and 1. This may be a fundamental property of nature or merely
an accidental property of the low energy
spectrum that we have been able to
observe so far. Indeed, fractionally charged particles may exist provided
they are sufficiently heavy or sufficiently rare.

In the superstring standard--like model the
following mass terms for the fractionally charged states are obtained
from nonrenormalizable terms,
$$\eqalignno{H_3H_4&(\Phi_{12}\Phi_2^+\Phi_1^+
+\Phi_{12}{\bar\Phi}_2^-{\bar\Phi}_1^-+\xi_3\Phi_2^+{\bar\Phi}_1^-
+\xi_3{\bar\Phi}_2^-\Phi_1^+)&(23a)\cr
H_7H_8&(\xi_1{\bar\Phi}_1^-\Phi_3^++\xi_1
{\bar\Phi}_3^-\phi_1^+)&(23b)\cr
H_{11}H_{12}&
(\xi_2\Phi_2^+{\bar\Phi}_3^-+\xi_2{\bar\Phi}_2^-\Phi_3^+)&(23c)\cr
H_{29}H_{30}&(\xi_3{{\partial W}_3\over{\partial\xi_3}}+
\Phi_{12}({\bar\Phi}^+_i{\bar\Phi}^-_i)+
{\bar\Phi}_{12}({\Phi}^+_i{\Phi}^-_i)).&(23d)\cr}$$
{}From Eq. (23b) and Eq. (23c)
we learn that $H_7H_8$ and $H_{11}H_{12}$ acquire a large mass
by the non vanishing VEV of the fields
$\{\xi_1,\xi_2,{\bar\Phi}_{1,2,3}^-,{\Phi}_{1,2}^+\}$.
Since  ${\bar\Phi}_{1,2,3}^-,{\Phi}_{1,2}^+$ obtain a Planck scale VEV
the mass scale of these fractionally charged singlets is determined
by the VEV of $\xi_1,\xi_2$.

The term $H_3H_4\Phi_1^+\Phi_2^+\Phi_{12}$
induces an effective mass term
$H_3H_4\Phi_{12}\left
({{\langle\Phi_1^+\rangle\langle\Phi_2^+\rangle}\over{M^2}}\right)$,
where
$M$, ${\langle\Phi_1^+\rangle}$ and ${\langle\Phi_2^+\rangle}$ are
$O(M_{Pl})$. This term will give a heavy
mass term to $H_3H_4$ by the VEV
of $\Phi_{12}$. According to the F and D flatness solution, this VEV
vanishes at the Planck scale,
and is constrained by the yet unknown mechanism
for supersymmetry breaking. Thus, $\Phi_{12}$ may obtain a VEV which is
still tolerated by the requirement of $N=1$ space--time supersymmetry,
giving a superheavy mass to $H_3H_4$, which is beyond the reach
of present accelerators. Similarly, the term
$H_{29}H_{30}{\bar\Phi}_i^+{\bar\Phi}_i^-\Phi_{12}$ induces an effective
mass term $H_{29}H_{30}{{\langle{\bar\Phi}_i^+{\bar\Phi}_i^-\rangle}\over
{M^2}}\Phi_{12}$. From Eq. (15a)
${\langle{\bar\Phi}_i^+{\bar\Phi}_i^-\rangle}$ is $O(M_{Pl}^2)$.
Therefore this term is an effective mass term for $H_{29}H_{30}$
by the VEV of $\Phi_{12}$.

This result illustrates that all the fractionally charged states
are expected to decouple from the low energy spectrum. Since all the
fractionally charged states appear in vector--like representations
this result is expected.
The exact mass scales
can only be determined by resolving the problem of supersymmetry
breaking in these models.

\vfill
\eject

\centerline{\bf 6.  Higgs mass matrix }

The light Higgs spectrum is determined by the massless eigenstates of the
doublet Higgs mass matrix. The doublet mass matrix consists of the terms
$h_i{\bar h}_j\langle\Phi^n\rangle$, and is defined by
$h_i(M_h)_{ij}{\bar h}_j$, $i,j=1,2,3,4$
where $h_i=(h_1,h_2,h_3,h_{45})$
and ${\bar h}_i=({\bar h}_1,{\bar h}_2,{\bar h}_3,{\bar h}_{45})$.
At the cubic level of the
superpotential the Higgs doublets mass matrix is given by,
$$M_h={\left(\matrix{0&0&{{\bar\Phi}_{13}}&0\cr
   0&0&{{\bar\Phi}_{23}}&0\cr
   {\Phi_{13}}&0&0&{\Phi_{45}}\cr
   0&0&0&0\cr}\right).}\eqno(24)$$

The matrix $M_h$ is diagonalized by $S{M_h}T^{\dagger}$ where $S$ and $T$
are two unitary matrices and $(SM_hT^\dagger)_{ij}=m_i\delta_{ij}$.
It follows that $SM{M^\dagger}S=TM{^\dagger}MT=\vert{m}\vert^2$.
The $h$ and ${\bar h}$
mass eigenstates are obtained by evaluating the
eigenvalues and eigenstates
of $MM{^\dagger}$ and $M{^\dagger}M$, respectively.
The mass eigenvalues are given by
$$m_h;m_{\bar h}=
(0,0,\Phi_{13}^2+\Phi_{45}^2,\Phi_{13}^2+\Phi_{23}^2).\eqno(25)$$
The $h$ mass eigenstates are given by
$$\eqalignno{{h^\prime}=&(0,0,0,1);&(26a)\cr
 & (-{{{\bar\Phi}_{23}}\over{{\bar\Phi}_{13}}},1,0,0);&(26b)\cr
    &({{{\bar\Phi}_{13}}\over{{\bar\Phi}_{13}}},1,0,0);&(26c)\cr
&(0,0,1,0),&(26d)\cr}$$
and the ${\bar h}$ mass eigenstates are given by
$$\eqalignno{{\bar h}^\prime=&(1,0,0,
-{{{\bar\Phi}_{13}}\over{{\bar\Phi}_{45}}});&(27a)\cr
&(0,1,0,0);&(27b)\cr
&(0,0,1,0);&(27c)\cr
&(1,0,0,{{{\bar\Phi}_{45}}\over{{\bar\Phi}_{13}}}).&(27d)\cr}$$
Equations (25), (26) and (27) show that at the cubic
level of the superpotential
there are two pairs of light Higgs states. The number of light Higgs
pairs is reduced by taking into account higher order terms in the
superpotential. For example at the quintic level we obtain
the following non vanishing terms
$$h_2{\bar h}_{45}\Phi_{45}H_{25}H_{26}{\hskip 1cm};{\hskip 1cm}
  {\bar h}_2h_{45}{\bar\Phi}_{45}H_{23}H_{27}\eqno(28a,b)$$
These additional terms reduce the number of light Higgs pairs
to one pair. For example, if
$\langle{H_{25}}\rangle\sim
\langle{H_{26}}\rangle\sim{10^{14}}GeV$, one of the light pairs
receives a mass of $O(10^{10}GeV)$. At order $N=7$ we obtain
additional terms which may make the extra pair massive without
breaking $U(1)_{Z^\prime}$.
The remaining light combinations depend on the specific
entries in the Higgs mass matrix which become non zero and is highly
model dependent. For example, if the $1{\bar2}$ entry in equation
$(24)$ is non zero, the two light Higgs eigenstates will consist of
$h_{45}$ and a combination of ${\bar h}_1$ and ${\bar h}_{45}$.
Below I assume that only one pair of Higgs doublets
remain light. However, I do not make a specific assumption as to what
are the exact light eigenstates, but rather assume that the light pairs
may contain any of the states that remain light at the cubic level.
The purpose in doing so is to try to learn general properties
of the light spectrum rather than details which depend on specific
choices of flat directions.
{}From equations (26) and (27) it follows
that $h_3$ and ${\bar h}_3$ do not
appear in the light eigenstates. Therefore the light eigenstates
may contain only $(h_1,h_2,h_{45})$
and $({\bar h}_1,{\bar h}_2,{\bar h}_{45})$. The absence of $h_3$
and ${\bar h}_3$ from the light eigenstates results in $G_3$ being
identified with the lightest generation. As I show in the next
section, the states in $G_3$ do not couple directly to the light
Higgs eigenstates. Therefore, diagonal mass terms for $G_3$ do not
appear up to $N=8$. Consequently, after diagonalization of the
mass matrices, the states in $G_3$ will be the largest component in
lightest generation states.

\bigskip
\centerline{\bf 7.  Fermion masses}

One of the most fundamental problems in high energy physics is
the origin
and hierarchy of the fermion masses.
In this respect the Standard Model, and
point field theories in general,
can only be considered as successful attempts
to parameterize the observed mass spectrum. Superstring theory gives
a unique framework to understand the fermion mass hierarchy in terms of
symmetries which are derived in specific models, unlike point field
theories where the symmetries have to be imposed by hand.
Therefore
it is important to examine the structure of the fermion mass
matrices in specific superstring
models [\SFMM].

The class of superstring
standard--like models is an especially restrictive class of models
in which the fermion mass spectrum can be examined. A unique property
of the standard--like models is the possible connection between the
requirement of a supersymmetric vacuum at the Planck scale, via the
DSW mechanism, and the heaviness of the top quark relative to the
lighter quarks and leptons. The only standard--like models
which admit a solution to the set of F and D constraints are models
in which only $+{2\over3}$ charged quarks obtain trilevel Yukawa
couplings. Application of the DSW mechanism leaves a trilevel mass term
only to the
top quark. The mass terms for the lighter
quarks and leptons must come from higher order, nonrenormalizable, terms.
These terms become effective mass terms
for the lighter quarks and leptons
by applying the DSW mechanism, and are naturally suppressed relative to
the trilevel top Yukawa coupling. A second property, unique to the
standard--like models, is the fact that the massless spectrum contains
only three light generations. There are no extra generations and
mirror generations which become superheavy at some high scale.
This property of the standard--like models eliminates the ambiguity
in the identification of the different generations that exist in
other realistic superstring models [\REVAMP,\ALR].

The top quark mass term is obtained from ${\lambda_t}u_1Q_1{\bar h}_1$.
At the quartic level there are no potential mass terms for the quarks
and leptons.
At the quintic level, the following mass terms are obtained
$$\eqalignno{&d_2Q_2h_{45}{\bar\Phi}_2^-\xi_1,{\hskip .2cm}
       e_2L_2h_{45}{\bar\Phi}_2^+\xi_1&(28a)\cr
      &d_1Q_1h_{45}{\Phi}_1^+\xi_2,{\hskip .2cm}
       e_1L_1h_{45}{\Phi}_1^-\xi_2&(28b)\cr
      &u_2Q_2({\bar h}_{45}\Phi_{45}{\bar\Phi}_{23}+
  {\bar h}_1{\bar\Phi}_i^+{\bar\Phi}_i^-)&(28c)\cr
      &u_1Q_1({\bar h}_{45}\Phi_{45}{\bar\Phi}_{13}+
  {\bar h}_2{\Phi}_i^+{\Phi}_i^-)&(28d)\cr
      &(u_2Q_2h_2+u_1Q_1h_1)
  {{\partial W}\over{\partial\xi}_3}.&(28e)\cr}$$
At this level potential mass terms for the heaviest down quark and
charged lepton are obtained, $d_1Q_1h_{45}{\Phi}_1^+\xi_2$,
        $e_1L_1h_{45}{\Phi}_1^-\xi_2$.
{}From the solution to the F and D constraints $\vert\Phi_1^+\vert=
\vert\Phi_1^-\vert$. Therefore, $\lambda_b=\lambda_\tau$
at the unification
scale. However, the VEV of $\xi_2$ is not determined
by the F and D
constraints and is left as free parameter.

The charm quark obtains a mass term from
$u_2Q_2{\bar h}_1({\bar\Phi}_1^+{\bar\Phi}_1^-)$.
The charm quark mass
is suppressed by
${{({\bar\Phi}_1^+{\bar\Phi}_1^-)}\over{M^2}}$
relative to the top quark mass.
The suppression factor is expected to be of
about two orders of magnitude.
If we take $\langle\xi_1\rangle{\not=}0$ at the unification scale,
$d_2Q_2h_{45}{\bar\Phi}_2^-\xi_1$ and $e_2L_2h_{45}{\bar\Phi}_2^+\xi_1$
can give mass terms to the strange quark
and to the muon lepton.
According to
Eq. (20),
${\bar\Phi}_2^-
=({\sqrt{7\over2}}{{1+{\sqrt{2}}}\over{60}})^{1\over2}$
and ${\bar\Phi}_2^+=0$. Therefore according to this solution only the
strange quark get mass from this term. A modified solution which includes
${\bar\Phi}_2^+{\not=}0$ will give a mass term to the muon lepton
as well.
At this
level the
states in $G_3$ do not receive any mass terms. Therefore, $G_3$ is
identified with the lightest generation.

At every increasing order of nonrenormalizable
terms the number of potential
non vanishing terms increases exponentially.
A search up to $N=8$ was
performed.
Several observations simplify the analysis. First, there is
no component of $h_3$ or ${\bar h}_3$ in the light Higgs
representations. Second, there are several scales in the model.
The leading scale correspond to the VEVs of singlets fields.
There are two non abelian hidden gauge groups
$SU(5)_H\times SU(3)_H$, with matter in fundamental representations
(see tables 2,3). These hidden gauge groups produce two additional
scales in the model, which correspond to the scales at which
their couplings become strong. I assume that $\Lambda_5>>\Lambda_3$.

At order $N=6$
all the up quark mass terms
are suppressed by  at least ${{\Lambda_3^2}\over{M^2}}$.
There are no diagonal mass terms for the states in $G_3$.
In the down quark and charged lepton sectors we obtain
the following leading terms,
$$\eqalignno{
&d_3Q_2h_{45}\Phi_{45}V_6V_9,{\hskip 2cm}
d_2Q_3h_{45}\Phi_{45}V_5V_{10},&(29a)\cr
&d_3Q_1h_{45}\Phi_{45}V_2V_9,{\hskip 2cm}
d_1Q_3h_{45}\Phi_{45}V_1V_{10},&(29b)\cr
&e_3L_2h_{45}\Phi_{45}V_8V_{11},{\hskip 2cm}
e_2L_3h_{45}\Phi_{45}V_7V_{12},&(29c)\cr
&e_3L_1h_{45}\Phi_{45}V_4V_{11},{\hskip 2cm}
e_1L_3h_{45}\Phi_{45}V_3V_{12}.&(29d)
\cr}$$
At order $N=6$ we obtain generational mixing in the down quark sector
and in the charged lepton sector. In the quark sector the mixing
is proportional to ${{\Lambda_{3_H}^2}\over{M^2}}$. In the leptonic
sector it is proportional to ${{\Lambda_{5_H}^2}\over{M^2}}$.
It may be possible,
(and desirable)
to reverse this result by changing some of
the generalized GSO phases.
The importance of this result is to show that
generational mixing is obtained.
The symmetry between the down quark
and charged lepton sectors is broken
as the relative
magnitude of the mixing is related
by ${\Lambda_{3_H}^2\over\Lambda_{5_H}^2}$.

At order $N=7$ we obtain the following leading terms.
In the down quark sector all the generational mixing terms
are  proportional to
 ${{\Lambda_{3_H}^2}\over{M^2}}$, and are a small correction to the sixth
 order terms. Similarly, there are small corrections, of order one
 percent (assuming ${{\Phi}\over{M}}\sim{1\over{10}}$)
 to the diagonal quintic order terms. There is no diagonal
 term of the form $d_3Q_3h$ or $e_3L_3h$.

 In the up quark sector we obtain non vanishing generation mixing
terms. Here I list only the leading terms which are proportional to
${{\Lambda_{5_H}^2}\over{M^2}}$.
$$\eqalignno{
&u_3Q_2{\bar h}_1\Phi_{45}{\bar\Phi}_2^-V_{7}V_{12}{\hskip .8cm}
u_2Q_3{\bar h}_1\Phi_{45}{\bar\Phi}_3^-V_{7}V_{12}&(29a)\cr
&u_3Q_1{\bar h}_1\Phi_{45}{\bar\Phi}_1^-V_{3}V_{12}{\hskip .8cm}
u_1Q_3{\bar h}_1\Phi_{45}{\bar\Phi}_3^-V_{3}V_{12}&(29b)\cr
&u_2Q_1{\bar h}_1\Phi_{45}{\bar\Phi}_1^-V_{3}V_{8}{\hskip .8cm}
u_1Q_2{\bar h}_1\Phi_{45}{\bar\Phi}_2^-V_{3}V_{12},&(29c)
\cr}$$
with additional terms, obtained by replacing ${\bar h}_1$ by
${\bar h}_2$.
At order $N=8$ generational mixing which is proportional to
${{\Lambda_{5_H}^2}\over{M^2}}$, appears in the down quark sector.
Up to level $N=8$ the diagonal mass terms for the states in $G_3$
are suppressed by at least $({{M_{SUSY}}\over{M_{Pl}}})^2$. Therefore,
these states are identified with the lightest generation states.
The identification of $G_3$ with the lightest generation is
unambiguous and completetly general. It is a general characteristic
of the class of standard--like models under consideration.
It follows from the structure of the boundary
condition vectors which characterizes these models.
At the level of the NAHE set there is a cyclic
symmetry between the vectors $b_1$, $b_2$ and $b_3$. Therefore,
there is universality among the generations. This cyclic symmetry
is broken by the vectors $\alpha$ and $\beta$. The vectors
$\alpha$ and $\beta$ are symmetric with respect to $b_1$ and $b_2$.
However, the cyclic symmetry between $b_1$, $b_2$ and $b_3$ is
broken. The universality among the three generations with respect
to the horizontal $U(1)$ symmetries is still unbroken. The symmetries
of the spin structure determine the allowed terms in the cubic
superpotential. These symmetries and the requirement of F flatness
impose $\langle\Phi_{12},{\bar\Phi}_{12},\xi_3\rangle\equiv0$.
Therefore, requiring D flatness by applying the DSW mechanism
removes the degeneracy among the generations and forces $h_3$ and
${\bar h}_3$ to become superheavy. Since the remaining
light doublets are not charged under $U(1)_3$, and because the only
$NS$ or $\alpha\beta$ fields with $\chi_{56}$ charge are $\phi_{12}$,
${\bar\Phi}_{12}$ and $\xi_3$, diagonal mass terms for the states
in $G_3$ are suppressed.

\bigskip
\centerline{\bf 8.  Conclusions }

In this paper I examined several aspects of nonrenormalizable terms
is a superstring derived standard--like model. This model belongs to a
class of standard--like models with unique characteristics. They
reproduce most of the properties of the Standard Model and
provide explanations to several fundamental puzzles beyond the Standard
Model. Among those, the replication of three and only three generations
of chiral fermions and the heaviness of the top quark relative to
the lighter quarks and leptons.

Nonrenormalizable terms play a pivotal role in the phenomenology of
these models. Due to nonrenormalizable terms the  preferred vacuum at the
Planck scale extends the Standard Model gauge symmetry by an additional,
generation independent, $U(1)$ symmetry. This $U(1)$ symmetry is
uniquely determined to be, $U(1)_{Z^\prime}={1\over2}U(1)_{B-L}-
{2\over3}U(1)_{T_{3_R}}$. Breaking of the gauge symmetry, at the
Planck scale, directly to the
Standard Model results in violation of the cubic level F flatness
solution or in induction of dimension four operators, which mediate
rapid proton decay. Thus, this model predicts the existence of an
additional neutral gauge boson below the Planck scale.
Nonrenormalizable terms lead to decoupling of the fractionally charged
states from the massless spectrum.

The most important function of
nonrenormalizable terms is in generating the hierarchy of the
fermion mass spectrum. This function of nonrenormalizable terms
is the fingers print of specific superstring models. The origin
of the fermion mass spectrum is perhaps the most fundamental
problem in physics. The ability of superstring models to generate
the observed spectrum is the real challenge facing these models.
The standard--like models have the advantage that they
explain the mass hierarchy of the top quark relative to the
lighter quarks and leptons.
In this paper I demonstrated that the superstring standard--like
model can in principle account for the observed spectrum,
including generational mixing. Resolution of the problem of
supersymmetry breaking in these models, better understanding
of the dynamics of the hidden sector, and explicit calculation
of the coefficients of the higher order terms, will improve
our ability to obtain quantitative estimates.
Resolving these problems will uniquely determine the singlets VEVs,
the hidden sector condensates, and the numerical coefficients of
the higher order terms. Thus, yielding a full quantitative
confrontation versus the low energy observations. I will come back
to the phenomenology extracted from these models in future publications.

\bigskip
\centerline{\bf Acknowledgments}
I would like to thank John Rizos for useful discussions and the
SSC and SLAC laboratories for their hospitality while part of this
work was conducted.
This work is supported in part by a Feinberg
School Fellowship.

\refout

\vfill
\eject

\bye